\documentclass[twoside,a4paper]{JHEP3}

\usepackage[dvips]{graphicx}
\usepackage{amsmath}
\usepackage{amsfonts}
\usepackage{cite}
\usepackage{epsfig}
\usepackage[english]{babel}
\usepackage[latin1]{inputenc}
\usepackage[T1]{fontenc}
\usepackage{float}
\usepackage{caption}
\usepackage[amssymb]{SIunits}
\usepackage{microtype}

\newcommand{\mc}{\mathcal}
\newcommand{\GeV}{\giga\electronvolt}
\newcommand{\TeV}{\tera\electronvolt}
\renewcommand{\Re}{\mathrm{Re}}

\DeclareMathOperator{\Vol}{Vol}

\newcommand{\IP}{\mathbb{P}}
\newcommand{\cV}{\mathcal{V}}
\newcommand{\cK}{\mathcal{K}}
\newcommand{\cO}{\mathcal{O}}

\title{SUSY Breaking in Local String/F-Theory Models}

\author{R.~Blumenhagen$^1$, J.\,P.~Conlon$^2$, S.~Krippendorf$^3$, S.~Moster$^1$, F.~Quevedo$^{3,4}$\\
$^{1}$ Max-Planck-Institut f\"ur Physik, F\"ohringer Ring 6, 80805 M\"unchen, Germany\\
$^{2}$ Rudolf Peierls Centre for Theoretical Physics, 1 Keble Road, Oxford OX1 3NP, UK\\
$^{3}$ DAMTP, University of Cambridge, Wilberforce Road, Cambridge, CB3 0WA, UK\\
$^{4}$ CERN PH-TH, CH 1211, Geneva 23, Switzerland.}


\preprint{%
MPP-2009-76\\
OUTP-09/14P\\
DAMTP-2009-48\\
CERN-PH-TH/2009-089%
}

\abstract{%
We investigate bulk moduli stabilisation and supersymmetry breaking in local
string/F-theory models where the Standard Model is supported on a del Pezzo
surface or singularity. Computing the gravity mediated soft terms on the
Standard Model brane induced by bulk supersymmetry breaking in the LARGE volume
scenario, we explicitly find suppressions by $M_s / M_P \sim \mc{V}^{-1/2}$
compared to $M_{3/2}$. This gives rise to several phenomenological scenarios,
depending on the strength of perturbative corrections to the effective action
and the source of de Sitter lifting, in which the soft terms are suppressed by
at least $M_P / \mc{V}^{3/2}$ and may be as small as $M_P / \mc{V}^2$.
Since the gravitino mass is of order $M_{3/2} \sim M_P / \mc{V}$, for \TeV\
soft terms all these scenarios give a very heavy gravitino ($M_{3/2} \geq
\unit{10^{8}}{\GeV}$) and generically the lightest moduli field is also heavy
enough ($m \geq \unit{10}{\TeV}$) to avoid the cosmological moduli problem. For
\TeV\ soft terms, these scenarios predict a minimal value of the volume to be
${\mc{V}} \sim 10^{6-7}$ in string units, which would give a unification scale
of order $M_{GUT} \sim M_s \mc{V}^{1/6} \sim \unit{10^{16}}{\GeV}$. The strong
suppression of gravity mediated soft terms could also possibly allow a scenario
of dominant gauge mediation in the visible sector but with a very heavy
gravitino $M_{3/2} > \unit{1}{\TeV}$.}

\begin{document}

\clearpage

\section{Introduction}

During the last years moduli stabilisation, in particular for Type~IIB
orientifolds on compact Calabi--Yau threefolds, has been under intense study and
several scenarios have been proposed. The original example is the KKLT
scenario~\cite{Kachru:2003aw}, where dilaton and complex structure moduli are
fixed at tree-level by fluxes while K\"ahler moduli are stabilised via
instanton generated terms in the superpotential
\begin{equation}
  W=W_0+A\, e^{-a\, T_s} \;.
\end{equation}
As a generalisation of this, including also next-to-leading order corrections to
the K\"ahler potential, the volume of the compactification manifold $\mc{V}$ can
be stabilised at exponentially large values~\cite{Balasubramanian:2005zx,
Conlon:2005ki}. These large volume minima are quite generic~\cite{Cicoli:2008va}
and exist whenever there is a four-cycle in the Calabi--Yau threefold which is
shrinkable to zero size, such that the total volume of the space remains finite.
These supersymmetry breaking Type~IIB vacua have been called the LARGE Volume
Scenario (LVS). Their phenomenological features were studied in very much detail
for the string scale in the intermediate regime $M_s \simeq
\unit{10^{11}}{\GeV}$~\cite{Conlon:2006wz,Conlon:2007xv} leading to \TeV\ soft
terms and intermediate scales for the neutrino and axion sector of the MSSM in
the preferred range. For computing the high scale soft terms it was assumed that
the D7-branes supporting the MSSM gauge and matter fields wrap the same
four-cycle supporting also the D3-brane instanton. In~\cite{Blumenhagen:2007sm},
it was pointed out that instanton zero mode counting actually forbids such a
scenario and that the D7-branes and instantonic D3-branes should better wrap
distinct four-cycles in the underlying Calabi--Yau manifold. In fact the
K\"ahler moduli associated to the sizes of the four-cycles wrapped by the
D7-branes can be stabilised by D-terms, often at string scale size at the
boundary of the K\"ahler cone. In this sense the MSSM branes are sequestered
from the bulk of the Calabi--Yau.

A way of realising the MSSM gauge and matter fields in the LVS is by studying
fractional D3-branes at the singular point, discussed in~\cite{aiqu,
Conlon:2008wa}. Various low-energy models were studied on the first two del
Pezzo surfaces $dP_0$ and $dP_1$, allowing for both GUT-like and extended MSSM
scenarios. From the effective field theory point of view, both scenarios are
similar since after stabilising the moduli, both vacua are at or close to the
singular point.

A very similar scenario was proposed recently in the context of local F-theory
models with an $SU(5)$ GUT brane. First of all, it was realised that some of the
model building problems one had with realising simple GUT groups in orientifold
constructions~\cite{Blumenhagen:2001te} are nicely reconciled in F-theory models
on elliptically fibered Calabi--Yau fourfolds~\cite{Donagi:2008ca,
Beasley:2008dc, Beasley:2008kw, Donagi:2008kj, Donagi:2009ra, Marsano:2009ym,
Collinucci:2009uh, Blumenhagen:2009up}. The reason for this substantial
improvement is that F-theory is genuinely non-perturbative and also allows for
the appearance of exceptional groups $E_6$, $E_7$, $E_8$, which are supported
along a complex surface in the base threefold, over which the elliptic fiber
degenerates appropriately.\footnote{In the Type~IIB interpretation, such loci
support general $(p,q)$ seven brane systems, where the extra states are realised
by massless string junctions.} As a consequence, by a further breaking also the
spinor representation of an $SO(10)$ GUT and the top-quark Yukawa couplings
$10\, 10\, 5_H$ in $SU(5)$ GUTs can be realised.

Moreover, it was proposed in~\cite{Beasley:2008kw} that such models can allow
for an essentially local treatment, if there exists a limit in which gravity
decouples from the gauge theory on the GUT brane. Geometrically this means that
the space transverse to the brane can become arbitrary large or from a different
perspective that the four-cycle the brane is wrapping can shrink to zero size.
Such four-cycles are so-called del Pezzo surfaces, which are $\IP^2$ blown up at
up to eight points. Since there exists a limit where gravity decouples from the
physics on the $SU(5)$ brane, one expects that for gravity induced couplings on
the brane there exists an $M_{s}/M_P$ suppression relative to their general
values. This decoupling argument was used heavily when studying the further
phenomenological implications of local F-theory GUTs~\cite{Heckman:2008es,
Heckman:2008qt}. In particular, it was argued that, since gravity/moduli
mediated supersymmetry breaking soft terms on the GUT brane should also
experience such a suppression, gauge mediation could become the dominant source.
Indeed, under this assumption a very nice numerology for the soft terms was
deduced, which besides flavour universality includes a solution to the $\mu /
\mu B$ problem and a candidate for the QCD axion solving the strong CP-problem.
The gauge mediation was parametrised in the usual way by the non-zero VEV of a
scalar field $\langle X\rangle = x + \theta^2 F$ mediated to the MSSM by charged
messenger fields. Note that in this scenario the gravitino mass was assumed to
be dominantly set by gauge mediation. Therefore, the gravitino was the LSP with
a mass of $\unit{1}{\GeV}$.

Since both the LARGE volume scenario and local F-theory GUTs require the same
kind of geometry, i.\,e.\ Calabi--Yau respectively base threefolds containing
del Pezzo surfaces, it is natural to combine these two set-ups and study, for
this concrete moduli stabilisation mechanism in the bulk, the computable effects
of gravity mediation for the physics on the GUT brane. It is the primary aim of
this paper, to compute for a minimal set-up these gravity mediated soft terms
explicitly and compare them with the expectation of an $\mc{V}^{-\frac{1}{2}} =
M_s / M_P$ suppression. Indeed, as we will show such a computation requires to
compute the soft terms at next-to-leading orders in $1/\mc{V}$.

This paper is organised as follows: In section~2 we review the geometric
framework of local GUTs and the LARGE volume approach to moduli stabilisation
that is applicable in this regime. In section~3 we describe the computation of
gravity mediated soft terms. We describe how the soft terms cancel at
$\mc{O}(M_{3/2})$ and how it is necessary to consider sub-leading corrections.
We find sub-leading contributions to soft terms at order $\mc{O}(M_{3/2} /
\sqrt{\mc{V}}) = \mc{O} \left(\frac{M_{3/2}^{3/2}}{M_P^{1/2}} \right)$. In
certain circumstances these contributions can also cancel and we give a set of
well posed assumptions when this can occur. In section~4 we discuss the
implications from these soft terms for both gauge mediation and the cosmological
moduli problem, and in section~5 we conclude.

\section{Effective field theories and moduli stabilisation}

The minimal set-up we are investigating in this paper is that we have a
threefold with at least three four-cycles, one large cycle and two small del
Pezzo four-cycles, i.\,e.\ the threefold is of the (strong) swiss-cheese type.
One of the del Pezzos supports the $SU(5)$/MSSM gauge theory while the other one
can support a D3-brane instanton inducing a non-perturbative contribution to the
superpotential. Therefore, for the size of the overall volume and the instanton
four-cycle (without any further contributions) there exists the
non-supersymmetric AdS-type LARGE volume minimum. Since the GUT brane is
localised on a del Pezzo surface orthogonal to the instantonic del Pezzo and the
size of the GUT brane is fixed by D-terms at small values, the previous
computations of the gravity induced soft terms should be modified. The same
calculation is also necessary for the case that the GUT cycle is collapsed at
the quiver locus.

There are two basic regimes where the effective field theory (EFT) for light
modes is reliable:
\begin{itemize}
  \item{} All of the 4-cycles, including the standard model or GUT cycle are
          larger than the string scale. This is the geometric regime.
  \item{} The size of the standard model cycle is much smaller than the string
          scale. It is a standard blow-up mode expanded around its vanishing
          value corresponding to the del Pezzo singularity. Fortunately string
          theory is under control at the singularity and the EFT can be safely
          defined in an expansion on the blow-up mode.
\end{itemize}

Since the D-term conditions tend to prefer a small value of the standard model
cycle, it is important to understand the physics in both regimes of validity of
EFT\@. It is clear that these are two different effective field theories for
standard model physics. But, as we will see, since the standard model cycle does
not participate in the breaking of supersymmetry, the structure of soft breaking
terms will be the same in both cases.

Let us discuss the ingredients in some more detail.

\subsection{Gauge couplings on the GUT brane}

Let us recall the set-up for  Type~IIB respectively F-theory GUT models, where
we use for concreteness the Type~IIB orientifold language
of~\cite{Blumenhagen:2008zz, Blumenhagen:2008aw}. We consider the Type~IIB
string compactified on a compact Calabi--Yau threefold $\mc{M}$ modded out by an
orientifold projection $\Omega\, \sigma\, (-1)^{F_L}$. The holomorphic
involution is such that one gets O7- and O3-planes. The base of the
corresponding elliptically fibered four-fold is then $B_3=\mc{M}/\sigma$. The
$SU(5)$ GUT is localised on D7-branes wrapping a rigid del Pezzo surface $D_a$.
The resulting tree-level $SU(5)$ gauge kinetic function $f_{SU(5)} =
\frac{4\pi}{g^2_X} + i \Theta$ is simply given by
\begin{equation}
  f_{SU(5)} = T_a = \frac{1}{2\, g_s \ell_s^4} \int_{D_a} J \wedge J
      + i \int_{D_a} C_4\; ,
\end{equation}
where $g_s = e^\varphi$ denotes the string coupling constant and $\Vol(D_a) =
\frac{1}{2} \int_{D_a} J \wedge J$ is the volume of the del Pezzo surface $D_a$.

In orientifold models, we actually get on a stack of five D7-branes the
Chan--Paton gauge group $U(5)$, which allows for a non-vanishing gauge flux
$\mc{F}_a$ in the diagonal $U(1) \subset U(5)$. Since a del Pezzo is rigid and
does not even contain any discrete Wilson lines, the gauge symmetry is broken to
$SU(3) \times SU(2) \times U(1)_Y$ by a non-trivial $U(1)_Y$ gauge flux
$\mc{F}_Y$ supported on a two-cycle $C_a \in H_2(D_a,\mathbb{Z})$ which is
trivial in $H_2(\mc{M},\mathbb{Z})$~\cite{Beasley:2008kw, Donagi:2008kj}. As
explained in~\cite{Blumenhagen:2008aw}, this way of breaking the $SU(5)$ gauge
group leads to a specific pattern of MSSM gauge couplings at the unification
scale
\begin{equation}
\label{gaugekin}
  f_i = T_a - \frac{1}{2} \, \kappa_i \, S, \qquad i \in \{1,2,3\} \; ,
\end{equation}
with
\begin{eqnarray}
  \kappa_3 &=& \int_{D_a} \mc{F}^2_a,  \qquad \kappa_2 = \int_{D_a} \mc{F}^2_a
       + \mc{F}^2_Y + 2 \, \mc{F}_a\, \mc{F}_Y \\
  \kappa_1 &=& \int_{D_a} \mc{F}^2_a + \tfrac{3}{5} (\mc{F}^2_Y + 2\,
       \mc{F}_a\, \mc{F}_Y) \; .
\end{eqnarray}
For concreteness we are using these orientifold relations in the following.

In the limit that the cycle is collapsed to the singularity, the gauge kinetic
function takes a similar form:
\begin{equation}
  f_i = \delta_i S + s_{ik} T_k \; ,
\end{equation}
where now $T_k$ has to be understood as the blow-up modes that resolve the
singularity. For $\mathbb{Z}_n$ singularities $\delta_i$ is universal; however
for more complicated singularities $\delta_i$ can be non-universal. For
applications to unification, we are interested in singularities where the
different gauge groups have universal couplings at the singularity.

For both classes of local models the GUT unification scale and string scale
differ significantly by a factor of the bulk radius. More precisely, the GUT
unification scale $M_X$ is given by $M_X = R M_s$, where $R \sim \mc{V}^{1/6}$
is the bulk radius of the Calabi--Yau in string units. This can be seen through
the Kaplunovsky--Louis relation between physical and holomorphic gauge
couplings,
\begin{equation}
  \begin{split}
  \label{KL}
   g_a^{-2} (\Phi,\bar{\Phi}, \mu) &= \Re(f_a(\Phi)) + \frac{\left(
      \sum_r n_r T_a(r) - 3 T_a(G)\right)}{8 \pi^2} \ln \left(
      \frac{M_P}{\mu}\right) + \frac{T(G)}{8 \pi^2} \ln g_a^{-2}
      (\Phi, \bar{\Phi}, \mu) \\
  & + \frac{(\sum_r n_r T_a(r) - T(G))}{16 \pi^2} \hat{K}(\Phi, \bar{\Phi})
    - \sum_r \frac{T_a(r)}{8 \pi^2} \ln \det Z^r(\Phi, \bar{\Phi}, \mu) \; .
  \end{split}
\end{equation}
Using the IIB K\"ahler potential $\hat{K} = - 2 \ln \mc{V}$ and the behaviour
for local models $\hat{Z} = \mc{V}^{-2/3}$ we obtain
\begin{equation}
\label{mirage}
  g_a^{-2}(\mu) -\frac{T(G)}{8 \pi^2} \ln g_a^{-2}(\mu) = \Re(f_a(\Phi))
     + \beta_a \ln \left( \frac{(RM_s)^2}{\mu^2} \right) \; ,
\end{equation}
giving effective unification at $R M_s$. As described in~\cite{Conlon:2009xf,
09061920, JCPalti}, at the string level this dependence arises from the presence
of tadpoles that are sourced in the local model but are only cancelled globally.
This comes from the fact that the $U(1)_Y$ flux that breaks the GUT group is on
a two-cycle that is non-trivial in $H_2(D_a, \mathbb{Z})$ and trivial in
$H_2(\mc{M},\mathbb{Z})$. Locally the $U(1)_Y$ flux sources an RR tadpole, which
is in fact absent globally due to the triviality of the cycle. The finiteness of
threshold corrections is tied to the absence of RR tadpoles, but the triviality
of $C_a$ requires knowledge of the global geometry, leading to the presence of
the scale $R M_s$.

\subsection{Moduli Stabilisation}

So far we essentially considered a local part of the overall Calabi--Yau
geometry where the GUT physics is localised. As has been pointed out
in~\cite{Beasley:2008kw}, supersymmetry breaking on a hidden D-brane and
mediation via gauge interaction to the visible GUT brane might also partly allow
a completely local treatment. This presumes of course that all possible Planck
scale suppressed terms are sub-leading.

In this paper, we do not postpone these global issues but instead continue the
quite successful investigation of moduli stabilisation in the framework of Type
IIB orientifolds. More concretely, we consider a set-up where the bulk moduli
orthogonal to $D_a$ are stabilised by the LARGE volume scenario and compute the
induced soft terms on the $SU(5)$ brane, respectively the del Pezzo singularity
constructions.

\subsubsection{Fixing the non-standard model/GUT cycles}

For self-consistency let us review briefly the main ingredients for the KKLT
respectively LARGE volume scenario.

At order $\mc{V}^{-2}$ in the large volume expansion, the complex structure
moduli and the dilaton are stabilised by a non-trivial $G_3$-flux giving rise to
a tree-level superpotential of the form~\cite{Gukov:1999ya}
\begin{equation}
  \label{w_GVW}
   W_{\text{flux}} = \int_{\mc{M}} G_3 \wedge \Omega_3 \; .
\end{equation}
The resulting scalar potential is of the no-scale type, with the K\"ahler moduli
still flat directions.

In the LARGE volume scenario the no-scale structure is broken by a combination
of $\alpha'$-corrections to the K\"ahler potential and a D3-instanton correction
to the superpotential. Concretely, the K\"ahler potential including
$\alpha'$-corrections~\cite{Becker:2002nn} reads
\begin{equation}
  \label{kahl-pot}
  \mc{K} = -2\ln \left( \mc{V} + \tfrac{\hat{\xi}}{2} \right)
      - \ln \left( S+\bar{S} \right)  + \mc{K}_{\text{CS}} \; ,
\end{equation}
where $\hat{\xi} = \xi / g_s^{3/2}$ and $g_s$ is the string coupling. The
resulting inverse K\"ahler metric for the K\"ahler moduli $T_a$ and the
axio-dilaton $S$ reads
\begin{equation}
  \begin{split}
    \mc{K}^{a \bar{b}} &= -2\,\left( \mc{V} + \tfrac{{\hat{\xi}}}{2} \right)
       \left( \frac{\partial^2 \mc{V}}{\partial\tau_a \, \partial\tau_b}
        \right)^{-1} + \tau_a \, \tau_b\, \frac{4\,\mc{V} - \hat{\xi}}
        {\mc{V} - \hat{\xi} } \; ,\\
    \mc{K}^{a \bar{S}} &= -\frac{3}{2} \, (S+\bar{S}) \,
       \frac{\hat{\xi}}{\cV - \hat{\xi}} \, \tau_a \; ,\\
    \mc{K}^{S \bar{S}} &= \frac{(S + \bar{S})^2}{4} \,
       \frac{4 \, \cV - \hat{\xi}}{\cV - \hat{\xi}} \; .
  \end{split}
\end{equation}
For a D3-instanton to generate a contribution to the superpotential it has to
have the right number of zero modes. In fact an $O(1)$ instanton wrapping a
rigid four-cycle, which does not intersect any four-cycle carrying D7-branes,
has the right number of zero modes to give a contribution
\begin{equation}
  \label{kklt}
  W = W_0 + A\, e^{-a\, T_s}
\end{equation}
to the superpotential. Here $W_0$ is the value of the GVW superpotential in the
minimum and we implicitly assumed that one can first integrate out the complex
structure moduli and the axio-dilaton multiplet. This is justified by observing
that the instanton induced scalar potential is at order $\cO(\mc{V}^{-3})$ in
the LARGE volume expansion and that the K\"ahler metric for the complex
structure and K\"ahler moduli has a factorised form (for discussions of
integrating out moduli in supergravity see~\cite{ShantaInt, Marta, Achucarro,
GallegoSerone}).

Working in the large cycle regime, in the simplest case, one chooses the volume
$\mc{V}$ of the internal space to be of \emph{swiss-cheese} form with three
K\"ahler moduli
\begin{equation}
  \label{vol1}
  \mc{V} =  \left( \eta_b \tau_b \right)^{3/2} - \left(\eta_s \tau_s
       \right)^{3/2} - \left(\eta_a \tau_a \right)^{3/2} \; .
\end{equation}
Here $\tau_b$ determines the size of the Calabi--Yau and the small four-cycle of
size $\tau_s$ is wrapped by the D3-brane instanton. The resulting F-term
potential at order $\cO(\mc{V}^{-3})$ reads
\begin{equation}
  \begin{split}
  \label{scapot}
  V_F &= e^{\cK} \left( \cK^{a\bar{b}} \, D_a W \, D_{\bar{b}} \bar{W}
      - 3 \bigl| W \bigr|^2 \right)  \\
      &=  \lambda \, \frac{(a A)^2 \, \sqrt{\tau_s} \: e^{-2a \tau_s}}{\mc{V}}
          - \mu\,\frac{a \bigl|A W_0\bigr| \: \tau_s \, e^{-a \tau_s}}{\mc{V}^2}
          + \nu\,\frac{\xi\: \bigl|W_0\bigr|^2}{g_s^{1/2} \: \mc{V}^3}
          + \ldots \; ,
  \end{split}
\end{equation}
with coefficients $\lambda =\frac{g_s}{2} \frac{8}{3\eta_s^{3/2}}$, $\mu =
2g_s$, $\nu = \frac{3}{8}$, featuring the LARGE volume AdS minimum at $\mc{V}
\sim e^{a\tau_s}$. More details of this minimum are collected in
appendix~\ref{appa}, which allows one to compute the value of the scalar
potential~\eqref{scapot} in this minimum to be
\begin{equation}
  \label{vacuumenergy}
   V_0=-\frac{3}{16 \, a \tau_s} \, \frac{\xi}{g_s^{3/2}} \,
       \frac{W_0^2}{\mc{V}^3} \,  M_P^2 \; .
\end{equation}
Clearly it is negative, but due to a cancellation of the leading order terms, it
contains an extra suppression by $(a \tau_s) \simeq \log(\mc{V})$. For a
realistic model this negative vacuum energy has to be uplifted to $V_0 \simeq
0$. For the computation of the gravity and anomaly mediated soft terms in
section~\ref{sec:grav}, we will start by neglecting the effects of uplifting,
but will then consider the contributions of the uplifting sector.

\subsubsection{Fixing the standard model/GUT cycle}

Coming back to the GUT brane, following the zero mode arguments
in~\cite{Blumenhagen:2007sm}, we assume that the GUT branes are wrapping a
four-cycle of size $\tau_a$ which is ``orthogonal'' to the instanton cycle. As
mentioned, in Type~IIB orientifolds we allow for an additional gauge flux
$\mc{F}_a$ in the diagonal $U(1)_a \subset U(5)$ perturbative Chan--Paton gauge
group. Vanishing of the ``Fayet--Iliopoulos'' $U(1)_a$ D-term constraint (at
order $\mc{V}^{-2}$)
\begin{equation}
  \label{FI_term}
  \int_{D_a} J \wedge \mc{F}_a = 0
\end{equation}
implies that $\tau_a \to 0$ so that one is driven to the quiver locus where
$\alpha'$-corrections cannot be ignored. In the EFT the
condition~\eqref{FI_term} is essentially that the field dependent FI-term
vanishes $\mc{K}_{T_a} = 0$. Using the K\"ahler potential in both the geometric
and quiver regimes, this condition show explicitly a dynamical preference for a
collapsed cycle $\tau_a \to 0$. The F-term of the field $T_a$ is of the form
\begin{equation}
  F_a = e^{\mc{K}/2} \left( W_{T_a} + W K_{T_a} \right) \; .
\end{equation}
Since the superpotential $W$ does not depend on the modulus $T_a$ and the D-term
condition implies $\mc{K}_{T_a} = 0$, we can see that this field does not break
supersymmetry, i.\,e.\ $F_a = 0$. Notice that this conclusion will not be
modified by including perturbative and non-perturbative corrections to the
K\"ahler potential since these corrections will equally modify the D- and
F-terms. Since $\tau_a=0$ one finds that also $F^a=0$, which is a very important
conclusion, as it indicates that the standard model is somehow sequestered from
the sources of supersymmetry breaking.

A loophole to this argument is that it implicitly assumes that the standard
model fields, charged under the corresponding $U(1)$, will not get a VEV\@.
Otherwise they would contribute to the D-terms and cancel the contribution from
the FI-term. Even though this is desirable phenomenologically to avoid a large
scale breaking of the standard model symmetries, such as colour, it should be
the outcome of a calculation. We illustrate in the appendix in a toy model that
this is actually the case as long as the soft scalar masses are not tachyonic.

A direct consequence is that the soft terms on the GUT brane can only be
generated at ``sub-leading'' order by $F^b$, $F^{s}$ and $F^S$, i.\,e.\ by
moduli which are sort of sequestered from the GUT brane.

\subsection{Including matter fields}

So far we have concentrated only on the EFT for moduli fields and their
stabilisation. In order to study soft-supersymmetry breaking we need to properly
introduce the matter field dependence in the EFTs in both the geometric and
singular cycle regimes. The important term to be included is the matter fields'
K\"ahler potential $\tilde{K} = Z_{\alpha\beta} \varphi_\alpha \varphi^*_\beta +
\cdots$ with $Z_{\alpha\beta}$ a function of the moduli fields.

At this state, only the dependence on $\tau_b$ and $\tau_s$ is relevant, as all
the other fields do not break supersymmetry (to leading order). $Z$ should only
depend on $\tau_b$, $S$ and the K\"ahler modulus of the GUT brane $\tau_a$, so
$Z = Z(\tau_b, \tau_a, S)$. The leading order expression for $Z$ was determined
in~\cite{Conlon:2006tj} with $Z \sim 1/\mc{V}^{2/3}$ (see also \cite{Aparicio:2008wh}) which applies to both
chiral matter at magnetised D7-branes and to the better understood fractional
D3-branes at singularities. Since the $\alpha'$-corrections to the K\"ahler
potential are crucial to determine the large volume vacuum, consistency requires
that these corrections should also be included in the matter field K\"ahler
potential.  Unfortunately these corrections are not known at present. However,
as in the tree-level case, we are mostly interested on their overall volume
dependence.

Let us parametrise the $\alpha'$-corrections by a so far unknown function $f$:
\begin{equation}
  Z_{\alpha} = \frac{k_{\alpha}}{\tau_b}\left(1 + f\left(
    \frac{\Re(S)}{\tau_b} \right) \right) \; .
\end{equation}
The dependence of $f$ on the variables can only be in the indicated way in order
to have the right power in $g_s$. Now consider  the next-to-leading order
correction in $\alpha'$ to the tree-level result, which, we claim, must be of
the form:
\begin{equation}
\label{kaehleralpha}
  Z_{\alpha} = \frac{k_{\alpha}}{\tau_b}
      \left(1 - \delta \, \left( \frac{\Re(S)}{\tau_b} \right)^{\frac{n}{2}}
      + \cdots \right) \; ,
\end{equation}
with $n = 1, 2, \ldots$ denoting the $(\alpha')^n$ order of this term. The
question now is at which order in $(\alpha')^n$ the first correction appears.
Since we are only interested in the correction which does not include $\tau_a$,
we can use a scaling argument like in~\cite{Conlon:2006tj}. Assuming that the
physical Yukawa couplings do not depend on the overall volume of the space and
taking into account the K\"ahler potential~\eqref{kahl-pot}, the leading order
correction to the K\"ahler metrics were shown to scale as $\frac{k_{\alpha}}
{\tau_b}$. Then it is expected that also at next-to-leading order the scalings
must match, which means that also the K\"ahler metrics are corrected at order
$(\alpha')^3$. This argument shows that $n = 3$ is the smallest expected
correction in~\eqref{kaehleralpha} and then $Z_{\alpha} =
\frac{k_\alpha}{\tau_b} \left(1 -\delta\, \left(\frac{\Re(S)}{\tau_b}
\right)^{3/2} \right)$.

\subsection{Summary of EFTs}
We can finally summarise the expressions for the EFTs we are using for the two
relevant regimes:
\begin{enumerate}
  \item{} In the geometric regime the EFT is determined by:
    \begin{eqnarray}
      \mc{K} &=& -2 \ln \left( \mc{V} + \tfrac{{\hat{\xi}}}{2} \right)
          - \ln \left( S+\bar{S} \right)
          + \mc{K}_{\text{CS}} + Z \varphi\varphi^* + \cdots \, ,\\
           W &=& W_{0} + Ae ^{-aT_s} + W_{\text{matter}} \, ,\\
         f_i &=& T_a - \frac{1}{2} \kappa_i S \, ,
    \end{eqnarray}
    where $\mc{V} = \left( \eta_b \tau_b \right)^{3/2} - \left(\eta_s
    \tau_s \right)^{3/2} - \left( \eta_a \tau_a \right)^{3/2}$ and
    $Z = k \left(1 - \delta \left( \Re(S) \right)^{3/2} / \mc{V} \right)
       / {\mc{V}^{2/3}}$.

  \item{} In the singular cycle (blow-up) regime there is a slight change in
    the standard model cycle dependence of $\mc{K}$:
    \begin{eqnarray}
      \mc{K} &=& -2 \ln \left( \mc{V} + \tfrac{{\hat{\xi}}}{2} \right)
           + \alpha \frac{\tau_a^2}{\mc{V}} - \ln \left( S + \bar{S} \right)
           + \mc{K}_{\text{CS}} + Z \varphi\varphi^* + \cdots \, ,\\
          W &=& W_{0} + A e^{-aT_s} + W_{\text{matter}} \, ,\\
          f &=& \delta_i S+ s_{ik} T_k \, ,
    \end{eqnarray}
    with now $\mc{V} = \left( \eta_b \tau_b \right)^{3/2}  - \left(\eta_s \tau_s
    \right)^{3/2}$ and $Z = \left(\beta - \delta / \mc{V} + \gamma\tau_a^m
    \right) / \mc{V}^{2/3}$ with $m > 0$.
\end{enumerate}
Since in  both cases the standard model/GUT cycle does not break supersymmetry,
the structure of soft breaking terms will be essentially the same.

\section{Gravity mediated soft terms}
\label{sec:grav}

As we have seen, the LARGE volume minimum of the scalar potential breaks
supersymmetry, so that this breaking induces soft supersymmetry breaking terms
on the GUT brane. There are two sources which are relevant here. First, there
are of course the gravity mediated soft terms. However, since the GUT brane is
sequestered from the non-supersymmetric bulk one might expect that anomaly
mediation is the leading order contribution. In this section we compute the
gravity mediated soft terms, i.\,e.\ the gaugino- and sfermion-masses as well as
the $\mu$-, A- and B-terms. Moreover, we compute the anomaly mediated gaugino
masses. Let us emphasise again that the scenario differs from the usual
intermediate scale LARGE volume scenario in that the string scale is much higher
(we assume $M_s \sim \unit{10^{15}}{\GeV}$ for consistency with unification at
$M_X \sim \unit{10^{16}}{\GeV}$), and that the GUT or MSSM branes are wrapping a
four-cycle completely sequestered from the four-cycles supporting D3-brane
instantons.

First we express the string scale $M_s=(\alpha')^{-1/2}$ in terms of the
Planck scale and the volume $\mc{V}$ of internal Calabi--Yau (in Einstein frame
and in units of $\ell_s=2\pi \sqrt{\alpha'}$)
\begin{equation}
  \label{masses_01}
  M_s = \frac{\sqrt{\pi} \, g_s^{1/4}}{\sqrt{\mc{V}}} \: M_P \; .
\end{equation}

Thus we obtain $M_s \simeq \unit{10^{15}}{\GeV}$ and $M_X \simeq \unit{1.2 \cdot
10^{16}}{\GeV}$ for $\mc{V}=\cO(10^6 - 10^7)$, a value large enough to trust the
$\mc{V}^{-1}$ expansion. Moreover, we immediately realise that the LARGE volume
expansion parameter is directly related to the local GUT expansion parameter,
i.\,e.\ $\mc{V}^{-1/2} \simeq M_s / M_P$. For computing the gravitino mass we
simply utilise the general formula $M_{3/2} = e^{\frac{\mc{K}}{2}} \, W$ leading
in our case to
\begin{equation}
  \label{masses_grav}
  M_{3/2} = \frac{g_s^{1/2} \lvert W_0\rvert}{\sqrt{2} \, \mc{V}} \, M_P \; .
\end{equation}

\subsection{Structure of soft terms}

We are now in a position to compute each of the gravity mediated soft
supersymmetry breaking terms in this class of scenarios.

\subsubsection{Gaugino masses}

For gravity mediated supersymmetry breaking, the gaugino masses are calculated
as
\begin{equation}
  \label{gaugino_grav}
  M_{\widetilde G} = \frac{1}{2\, \Re(f_i)} \: F^{I} \, \partial_I f_i
\end{equation}
for $i=3, 2, 1$, where for the gauge kinetic functions we use~\eqref{gaugekin}
with $\tau_a \simeq 0$ due to the D-term constraint.

Since the GUT brane is sequestered from the bulk we have $F^a = 0$ and the only
contribution can come from the dilaton F-term $F^S = e^{\mc{K}/2}
\mc{K}^{S\bar{J}} \bar{F}_{\bar{J}}$. We assume that the F-term condition for the
axio-dilaton $F_S = 0$ is fulfilled at leading order. At next-to-leading order,
there are then only sub-leading contributions from $F_S$ as well as terms from
$F_b$ adding up to $F^S = \frac{3}{2\sqrt{2}} \, \gamma \, \frac{\xi}{g_s^2} \,
\frac{W_0}{\mc{V}^2}$ where $\gamma$ is an $\mc{O}(1)$ factor (see in the
appendix for a more detailed derivation). Thus, the gravity mediation induced
term for the gaugino masses reads:
\begin{equation}
 \label{gaugino_mass}
 M_{\widetilde G} = \frac{3}{4\sqrt{2}} \, \gamma \, \frac{\xi}{g_s} \,
    \frac{ \lvert W_0\rvert M_P }{\mc{V}^2}
     = \frac{3}{4} \, \gamma \, \frac{\xi}{g_s^{3/2}} \,
       \frac{M_{3/2}}{\mc{V}} \; ,
\end{equation}
independent of the MSSM gauge group factor, as the factor $\kappa_i$
in~\eqref{gaugekin} cancels. Here we have assumed that the D-term fixes the size
of the GUT four-cycle at small volume in string units, so that the leading
contribution to $\Re(f_i) \simeq 25$ comes from the gauge flux induced
correction $\simeq \kappa_i \Re(S)$.

\subsubsection{Squark/Slepton masses}

The scalar masses obtained for gravity mediation of supersymmetry breaking read
\begin{equation}
  \label{scalar-mass}
  M_{\widetilde{Q}}^2 =  M_{3/2}^2 + V_0 - F^I F^{\bar{J}} \partial_I
     \partial_{\bar{J}} \ln Z_{\alpha}  \; ,
\end{equation}
where the potential in the minimum $V_0$ is assumed to be already uplifted so
that $V_0 \simeq 0$.

Computing now the soft-sfermion masses, let us first discuss the tree-level term
in $Z_{\alpha}$. In this case~\eqref{scalar-mass} reduces to
\begin{equation}
   M_{\widetilde Q}^2 = M_{3/2}^2 - \frac{(F^b)^2}{4\tau_b^2} \; ,
\end{equation}
where we have neglected the vacuum energy in the minimum. Again, there is a
cancellation of the gravitino mass squared with the leading term in $(F^b)^2$.
The term quadratic in $F^S$ is sub-leading as being of order $\mc{V}^{-4}$. In
the appendix~\ref{appa} we compute the next-to-leading order term in $F^b$,
which reads
\begin{equation}
  (F^b)^2 \approx 4\tau_b^2 \bigg[ M_{3/2}^2 + \frac{3}{8a\tau_s}\,
     \frac{\xi}{g_s^{3/2}} \, \left(1+\frac{3}{2a\tau_s} \right) \,
      \frac{M_{3/2}^2}{\mc{V}} \bigg] \; .
\end{equation}
Therefore, one gets for the soft sfermion masses squared
\begin{equation}
  \label{sferma}
  \begin{split}
     M_{\widetilde Q}^2  &= - \frac{3}{16a\tau_s} \, \frac{\xi}{g_s^{1/2}}
       \, \left(1+\frac{3}{2a\tau_s}\right) \, \frac{\lvert W_0 \rvert^2
       M_P^2}{\mc{V}^3} \\
      &= - \frac{3}{8a\tau_s}\, \frac{\xi}{g_s^{3/2}}\,
        \frac{M_{3/2}^2}{\mc{V}} \; ,
 \end{split}
\end{equation}
which at this stage come out tachyonic.

Next we need to discuss the higher $\alpha'$-corrections
in~\eqref{kaehleralpha}. The term with the highest power in $1 / \mc{V}$ is the
one with $(F^b)^2 \partial_b \partial_b \log \cdots$. It is straightforward,
that for $\tau_a / \tau_b\ll 1$ this simplifies to
\begin{equation}
  \label{kaehlcora}
  F^m F^n \partial_m \partial_n \log \left(1 -
      \delta \left( \frac{\Re S}{\tau_b}\right)^{\frac{n}{2}} + \cdots \right)
    \simeq F^b\, F^b\, \frac{\delta n\, (n+2)\, (\Re S)^{\frac{n}{2}}}
         {4 \, \tau_b^{\frac{n}{2}+2}}
    \sim \frac{\delta}{g_s^{\frac{n-2}{2}}} \,
      \frac{|W_0 |^2 M_P^2}{\cV^{ \left(2 + \frac{n}{3}\right) }} \; .
\end{equation}
Therefore, if there would be corrections of order $n=1,2$, they would dominate
over the corrections in~\eqref{sferma}. It is precisely the third order
corrections in $\alpha'$ which contribute to the sfermion masses at the same
order in $1 / \cV$. Including also the other moduli fields in~\eqref{kaehlcora},
the overall value of the squared scalar masses will then be proportional to
$\delta - \xi/3$:
\begin{equation}
  \label{eqeq}
  M_{\widetilde {Q}}^2 = M_{3/2}^2 \left( -\frac{1}{4a\tau_s} \,
      \frac{\xi}{g_s^{3/2}\cV} + \frac{15(\delta - \xi/3)}
      {4 g_s^{3/2} \cV} \right) \; .
\end{equation}

Therefore, depending on the relative size of these two contributions one can get
tachyonic or non-tachyonic sfermion masses. Moreover, it also shows that for
$\delta = \xi/3$ there are further cancellations taking place at this order.
This is precisely the value one expects from the above mentioned scaling
argument of the physical Yukawa couplings.
Later we will give an argument under which quite general assumptions such
cancellations should occur. One of the assumptions will be that really the
uplifting sector is correctly taken into account, which leads to a further
dependence of the K\"ahler metric on a supersymmetry breaking field. Note that
indeed the soft sfermion masses~\eqref{sferma} are of the same order as the AdS
vacuum energy~\eqref{vacuumenergy}, indicating that in these computations the
uplift sector cannot be neglected.

\subsubsection{$\hat{\mu}/\hat{\mu} B$-terms}

The formula for the $\hat{\mu}$-term is
\begin{equation}
  \label{mu_term}
  \hat{\mu} = \left( e^{\mc{K} / 2} \mu + M_{3/2} Z
     - \bar{F}^{\bar{I}} \partial_{\bar{I}} Z \right) \,
      \left( Z_{H_1}Z_{H_2} \right)^{-1/2} \; ,
\end{equation}
where $\mu$ denotes the supersymmetric $\mu$-parameter, which we keep for
completeness, although it can be argued to vanish under very general
assumptions~\cite{Conlon:2006tj}. We assume again the K\"ahler
metric~\eqref{kaehleralpha} for the Higgs fields as well as for $Z$. Here again,
a cancellation of the second and the third term occurs. Note, if the $\mu$
parameter is not equal to zero, it dominates over the sub-leading terms stemming
from $F^b$. Dropping the factors of order one, we are left with:
\begin{equation}
   \hat{\mu} \approx \frac{\sqrt{g_s}}{\sqrt{2}} \frac{\tau_b}{\mc{V}} \, \mu
       - \frac{M_{\widetilde{G}}}{4a\tau_s} \; .
\end{equation}

The expression for $B\hat{\mu}$ is more complicated:
\begin{equation}
  \label{ugliest_formula_of_the_world}
  \begin{split}
    B\hat{\mu} = &( Z_{H_1}Z_{H_2})^{-1/2}
        \bigg( e^{\cK/2} \mu (F^I \partial_I \cK
        + F^I \partial_I \log \mu - F^I \partial_I
         \log (Z_{H_1}Z_{H_2}) - M_{3/2} ) \\
     &+ (2M_{3/2}^2 + V_0)Z - M_{3/2}\bar{F}^{\bar{I}} \partial_{\bar{I}} Z
        + M_{3/2} F^I (\partial_I Z - Z \partial_I
          \log (Z_{H_1}Z_{H_2})) \\
     &- F^{\bar{I}}F^{J}\left(\partial_{\bar{I}} \partial_J Z
        - (\partial_{\bar{I}}Z) \partial_J
         \log (Z_{H_1}Z_{H_2}) \right) \bigg) \; .
  \end{split}
\end{equation}
However, due to the simple K\"ahler metric and assuming that $\mu$ is just an
input parameter without any moduli dependence, after a long but straightforward
calculation, the result is rather simple:
\begin{equation}
  B\hat{\mu} = - \left( \frac{\sqrt{g_s}}{\sqrt{2}} \frac{\tau_b}{\cV} \mu +
 \frac{M_{3/2}}{2a\tau_s}  \right) M_{\widetilde{G}} \; ,
\end{equation}
where we have dropped again the order one constants $k_{H_i}$ and $z$.

\subsubsection{A-terms}
The A-terms are given by:
\begin{equation}
  A_{\alpha\beta\gamma} = F^I (\partial_I \cK) + F^I \partial_I
     \log Y_{\alpha\beta\gamma} - F^I \partial_I
       \log Z_{\alpha} Z_{\beta} Z_{\gamma} \; .
\end{equation}
The Peccei--Quinn shift-symmetry forbids a dependence of the holomorphic
superpotential on the axio-dilaton or K\"ahler moduli, thus the Yukawa couplings
$Y_{\alpha\beta\gamma}$ can only depend on the complex structure moduli and they
drop out.

There is a cancellation of $F^b$ in the remaining two sums and we are left with
\begin{equation}
  A_{\alpha\beta\gamma} = F^s (\partial_s \cK) + F^S (\partial_S \cK) \; .
\end{equation}
As listed in the appendix, $F^s (\partial_s \cK)$ is suppressed with respect to
$F^S (\partial_S \cK)$ by a factor of $1/a \tau_s$. As we are interested only in
orders of magnitude, we keep only the latter term and get as result:
\begin{equation}
  A_{\alpha\beta\gamma} \approx F^S (\partial_S \cK)
     = -\frac{3}{4\sqrt{2}} \, \frac{\xi}{g_s} \,
         \frac{\lvert W_0 \rvert}{\cV^2} \, M_P
     =  - M_{\widetilde{G}} \; .
\end{equation}

\subsubsection{Anomaly mediated gaugino masses}

Let us also now estimate the anomaly mediated gaugino mass. It is clear that,
for such a sequestered observable sector, one would have guessed that not
gravity mediation but anomaly mediation induces the leading order soft terms.
General formula for all the different soft terms are not available, so that in
this section we just compute the anomaly mediated gaugino masses. The expression
for them reads~\cite{Bagger:1999rd} (see
also~\cite{deAlwis:2008aq}):\footnote{We use a different sign convention for the
F-terms leading to a different sign in the second and third term in the anomaly
mediated mass term than in~\cite{Bagger:1999rd}.}
\begin{equation}
  \label{anom_gaugino}
  \begin{split}
    M_{\widetilde G}^{\text{anom}} = &-\frac{g^2}{16\pi^2} \bigg[
       (3T_G - T_R) M_{3/2} - (T_G - T_R) (\partial_I \cK) F^I \\
         &- \frac{2T_R}{d_R} F^I \partial_I \log \det Z_{\alpha\beta}
       \bigg] \; ,
  \end{split}
\end{equation}
where $T_G$ is the Dynkin index of the adjoint representation, normalised to $N$
for $SU(N)$, and $T_R$ is the Dynkin index associated with the representation
$R$ of dimension $d_R$, normalised to $1/2$ for the $SU(N)$ fundamental.

A careful calculation of $F^b$, worked out in the appendix, reveals that it is
proportional to the gravitino mass at leading order: $F^b \approx - 2\tau_b
M_{3/2} - \frac{\tau_b}{2a\tau_s}M_{\widetilde G}$. This leads to a precise
cancellation of $M_{3/2}$ in~\eqref{anom_gaugino}. The final expression for the
anomaly mediated gaugino mass for a $SU(N)$ gauge group is thus
\begin{equation}
  M_{\widetilde G}^{\text{anom}} = -\frac{g^2}{16\pi^2}
     \left[ \left( N - \tfrac{1}{2} \right) - \frac{1}{4a\tau_s}
     \left (3N - \tfrac{1}{2}\right) \right] \, M_{\widetilde G} \; .
\end{equation}
Surprisingly, though the gravity mediated contribution to the gaugino mass is
suppressed with respect to the gravitino mass by a factor of $(M_X / M_P)^2$,
anomaly mediation is not the dominating source for the gaugino mass. It is
suppressed by the usual one-loop factor with respect to the gravity mediated
contribution. One expects a similar suppressed behaviour for the other soft
terms, so that anomaly mediation is even sub-leading to gravity mediation.

\subsection{Summary of gravity mediated soft masses}

In the above computation of soft terms we have seen that the leading terms
cancel and that we need to include higher order corrections in $\mc{V}^{-1}$.
Since this scale is directly correlated with $\zeta = M_s / M_P$, we can express
these gravity mediated soft terms in terms of the scales $M_{3/2}$ and $\zeta =
M_s / M_P$. The results are listed in table~\ref{tab_gut}, where we have set the
supersymmetric $\mu$ parameter to zero and estimated
\begin{equation}
  \sqrt{\frac{\pi}{3 \xi}} \simeq \sqrt{\frac{400}{\chi(\mc{M})}} \simeq 1, \qquad
     \text{and} \quad g_s \simeq 1 \; .
\end{equation}

\begin{table}[htbp]
  \renewcommand{\arraystretch}{1.5}
  \begin{center}
    \begin{tabular}{|c|c|}
      \hline\hline
      soft-term & scale\\
      \hline
      $M_{\widetilde{G}}$   & $\frac{1}{4} \, M_{3/2} \, \zeta^2$  \\[0.1cm]
      \hline
      $M_{\widetilde{Q}}^2$ & $\frac{1}{16\log \zeta} \, M^2_{3/2} \, \zeta^2$ \\[0.1cm]
      \hline
      $\hat{\mu}$-term      & $\frac{1}{8 \log\zeta} M_{\widetilde{G}}$ \\[0.1cm]
      \hline
      $B\hat{\mu}$-term     & $\frac{1}{4\log \zeta} M_{3/2} \, M_{\widetilde{G}}$ \\[0.1cm]
      \hline
      $A$-term              & $-M_{\widetilde{G}}$                      \\[0.1cm]
      \hline\hline
    \end{tabular}
    \caption{Classical gravity mediated soft terms for a na\"{\i}ve computation of
      soft terms. Here the expansion parameter is $\zeta = M_s / M_P$. We have
      assumed the supersymmetric $\mu$-term to vanish~\cite{Conlon:2006tj}.}
    \label{tab_gut}
  \end{center}
\end{table}
\noindent
All soft terms in table~\ref{tab_gut} are suppressed by $(M_s / M_P)^2$ relative
to the na\"{\i}ve expectation $M^n_{3/2}$ with $n=1,2$ depending on the
mass-dimension. This explicitly demonstrates that gravity effects from the bulk
are suppressed on the shrinkable GUT cycles, which is \emph{the} main assumption
of the local F-theory GUTs.

However, as seen in the text in certain cases there can be more cancellations
leading to even higher suppressions. Indeed so far we have neglected the uplift
sector, but have seen that the sfermion masses are actually of the same order of
magnitude as the uplift so that it should better not be neglected. We now
discuss under which well-posed assumptions further cancellations are present.

\subsection{Uplift and cancellations}
\label{sec:decoup}

In the last section we have computed the gravity induced soft terms on the GUT
brane. As we have explained, the computation relies on assumptions about the
expansions of the matter metrics at higher orders in $\alpha'$. While such
corrections must surely be present, it is difficult to know the precise form of
these corrections. We have explicitly seen for the sfermion masses that these
corrections contribute at the same order in $1 / \mc{V}$ as the next-to-leading
order contributions from $F^b$. Indeed, as seen in eq.~\eqref{eqeq} there can
potentially be further cancellations at this order. We have also computed the
soft terms under the assumption of $V_0 = 0$, but have not taken into account
the contribution of the supersymmetry breaking from the uplifting sector to the
soft terms. To consider these possibilities, let us argue in this section, how
one can arrive at quite general statements by making some well posed assumptions
and exploiting the consequences of using the supergravity formalism.

Recall that the physical Yukawas are given by
\begin{equation}
  \hat{Y}_{\alpha \beta \gamma} = e^{\mc{K}/2} \frac{Y_{\alpha \beta
      \gamma}}{\sqrt{Z_{\alpha} Z_{\beta} Z_{\gamma}}} \; .
\end{equation}
The shift-symmetries of the K\"ahler moduli imply that they do not appear
perturbatively in the superpotential Yukawa couplings $Y_{\alpha \beta \gamma}$.
Let us make the assumption that the physical Yukawas, being local renormalisable
couplings, do not depend on the fields breaking supersymmetry. This includes the
volume and also the hidden sector fields that are responsible for uplifting and
giving vanishing cosmological constant. We also assume pure F-term uplifting.

Such supersymmetry breaking fields appear in the overall K\"ahler potential, and
the constraints of holomorphy then imply that in order for the physical Yukawas
to be independent of such fields,
\begin{equation}
  Z_{\alpha} = e^{\mc{K}/3} \; .
\end{equation}
Note that this includes the tree-level behaviour of local matter fields,
$Z_{\alpha} \sim \frac{1}{\mc{V}^{2/3}} \sim \frac{1}{(T_b + \bar{T}_b)}$. In
this case it follows that
\begin{equation}
  \label{cancel}
  \begin{split}
    m_Q^2 &=  V_0 + M_{3/2}^2 - F^m \bar{F}^{\bar{n}} \partial_m
       \partial_{\bar{n}} \ln {Z}_{\alpha} \\
          &=  V_0 + M_{3/2}^2 - F^m \bar{F}^{\bar{n}}
          \frac{\mc{K}_{m \bar{n}}}{3}
           = \frac{2}{3} V_0 =  0 \; ,
  \end{split}
\end{equation}
for the case of vanishing cosmological constant.

The A-terms also vanish under this assumption. In this case the A-terms can be
most intuitively written as
\begin{equation}
  A_{\alpha \beta \gamma} Y_{\alpha \beta \gamma}
        = F^I \partial_I \hat{Y}_{\alpha \beta \gamma} \; ,
\end{equation}
with $\hat{Y}_{\alpha \beta \gamma}$ the physical Yukawa couplings. So it
immediately follows that if the physical Yukawa couplings do not depend on the
fields breaking supersymmetry, the A-terms all vanish.

The anomaly mediated contribution for gaugino masses gives
\begin{equation}
  \label{anomaly}
  \begin{split}
    M^{\text{anom}}_{\widetilde{G}} &=   \frac{b_a}{16 \pi^2} M_{3/2}
         - \frac{(\sum_r n_r T_a(r) - T(G))}{16 \pi^2} F^m \partial_m
         \mc{K}(\Phi, \bar{\Phi})  + \sum_r \frac{n_r T_a(r)}{8 \pi^2}
         F^m \partial_m \ln Z^r(\Phi, \bar{\Phi}) \\
      &= \frac{b_a}{16 \pi^2}M_{3/2} - \frac{(\sum_r n_r T_a(r) - 3T(G))}
            {16 \pi^2} \frac{F^m \partial_m \mc{K}(\Phi, \bar{\Phi})}{3} \\
      &= \frac{b_a}{16 \pi^2} \left( M_{3/2} - \frac{1}{3}
             F^m \partial_m \mc{K} \right) \; ,
  \end{split}
\end{equation}
where we have used ${Z} = e^{\mc{K}/3}$. The size of the anomaly mediated
contributions to gaugino masses then depends on the size of $M_{3/2} -
\frac{1}{3} F^m \partial_m \mc{K}$. The no-scale cancellation for $\tau_b$
implies the $\mc{O}(\mc{V}^{-1})$ terms cancel with non-vanishing terms at
$\mc{O}(\mc{V}^{-2})$. However~\eqref{anomaly} also includes the hidden
uplifting sector, which must have $\mc{K}_{\phi \bar{\phi}} F^{\phi}
F^{\bar{\phi}} \sim \frac{1}{\mc{V}^3}$ (in order to uplift the vacuum energy to
Minkowski). At this level we therefore cannot rule out that $F^{\phi}
\partial_{\phi} \mc{K} \sim \mc{V}^{-3/2}$, giving gaugino masses of order
$\frac{g^2}{16 \pi^2} \frac{1}{\mc{V}^{3/2}}$.

For the $\mu$-term, we obtain
\begin{equation}
  \hat{\mu} = e^{\mc{K}/6} \mu + (M_{3/2} - \frac{1}{3} F^I \partial_I
      \mc{K}) \; .
\end{equation}

For the B-term, we have (assuming no moduli dependence in $\mu$)
\begin{equation}
  \begin{split}
    (B \mu) &=  (Z_{H_1} Z_{H_2})^{-1/2} \Bigg( e^{\mc{K}/2} \mu
         \left( F^I \partial_I \mc{K} - F^I \partial_I \ln (Z_{H_1} Z_{H_2})
         - M_{3/2} \right)  \\
       & + (2 M_{3/2}^2 + V_0) Z - M_{3/2} F^{\bar{I}} \partial_{\bar{I}} Z
         + M_{3/2} F^I \left( \partial_I Z - Z \partial_I \ln \left( Z_{H_1}
             Z_{H_2} \right) \right) \\
       & - F^I F^{\bar{J}} \left( \partial_{\bar{I}} \partial_J Z -
         (\partial_{\bar{I}} Z) \partial_J \ln \left( Z_{H_1} Z_{H_2} \right)
          \right) \Bigg) \; .
  \end{split}
\end{equation}
If we take $Z = Z_{H_1} = Z_{H_2} = e^{\mc{K}/3}$ then we eventually obtain
\begin{equation}
  B \mu = e^{\mc{K}/6} \mu \left( \frac{1}{3} F^I \partial_I \mc{K}
      - M_{3/2} \right) + \left\vert \frac{1}{3} F^I \partial_I \mc{K}
      - M_{3/2} \right\vert^2 \; .
\end{equation}
This implies that the $\mu$- and B-terms involve the same expression as appeared
in the anomaly mediated expression~\eqref{anomaly} and that the $\mu$- and $B
\mu$-term are of the same order as required for successful electroweak symmetry
breaking.

As we do not currently know the form of $\alpha'$-corrections to the matter
metrics, we do not know whether the form $Z = e^{\mc{K}/3}$ is correct. However
it is a natural choice in the sense that it simply says that the physical Yukawa
couplings, being local, do not depend on the value of bulk fields. In the
context of the $\zeta(3) \chi(\mc{M}) \alpha'^3$-correction that entered the
moduli stabilisation, this is equivalent to the statement that physical Yukawa
couplings do not alter if you perform a conifold transition in the bulk (which
alters the Calabi--Yau Euler number).

The advantage of phrasing the computation in this way is that we can say that
moduli generate soft scalar masses to the extent to which the physical Yukawa
couplings depend on the moduli. While not straightforward, it is in principle
easier to compute the dependence of physical Yukawa couplings on the moduli.
String CFT computations give the directly physical couplings and therefore one
could analyse for certain local models (for example for a stack of D3-branes at
an orbifold singularity in a compact space) whether the physical couplings do
depend on the volume through a direct vertex operator string computation.

We can also use~\eqref{cancel} to compute the minimal value of the soft scalar
masses. The complete cancellation in~\eqref{cancel} arose from the assumption
that the physical Yukawa couplings has no dependence on all fields with non-zero
F-terms. However we know this statement is not true. The dilaton has an
irreducible F-term of $\mc{O}(\mc{V}^{-2})$ and enters the physical Yukawas.
This provides a minimal value for the scale of the physical Yukawa couplings.

\section{Consequences for Supersymmetry Breaking}
\label{sec:gauge}

In this previous section we have seen that both gravity and anomaly mediated
contributions to soft terms occur at levels far lower than na\"{\i}ve
expectation. This gives novel phenomenological consequences for various aspects
of supersymmetry breaking, which we now discuss.

\subsection{Gauge mediated scenarios}

In local F-theory models an interesting proposal was made assuming a model of
gauge mediation to dominate supersymmetry breaking in the observable sector.
This is very interesting since it incorporates the positive properties of gauge
mediation, such as positive squared scalar masses and  flavour universality and
yet address its problems, such as the $\mu/B\mu$ problem. This proposal though
requires the following implicit assumptions:
\begin{enumerate}
  \item{} The mechanism responsible for moduli stabilisation, which was not
    considered, fixes moduli at a high mass and decouples from supersymmetry
    breaking.

  \item{} Introduce a new matter sector that breaks supersymmetry dynamically
    and a set of messengers that communicate this breaking to the standard model
    fields.

  \item{} An anomalous $U(1)$ was proposed to communicate both sectors and
    address the $\mu/B\mu$ problem of gauge mediation. The anomalous $U(1)$ is
    naturally as heavy as the string scale but has low-energy implications
    after being integrated out.
\end{enumerate}

These conditions look at first sight too strong and unnatural. Achieving moduli
stabilisation without supersymmetry breaking and small cosmological constant is
a very strong assumption not realised in any of the moduli stabilisation
scenarios so far. It is known that a supersymmetric vacuum in supergravity, such
as in KKLT before the uplifting, is naturally anti de Sitter since in that case
the vacuum energy is $V_0 = -3 M_{3/2}^2 M_P^2$, which is very large unless the
superpotential is tuned in such a way that it almost vanishes at the
supersymmetric minimum. Also a positive cosmological constant has to be induced
after supersymmetry breaking. If the local supersymmetry breaking is responsible
for this lifting then its effect should not have been neglected for moduli
stabilisation in the first place. Finally, it is not consistent to consider the
low-energy effects of a very heavy anomalous $U(1)$ without also including the
effects of the moduli fields which are generically much lighter than the string
or compactification scale. In particular the Fayet--Iliopoulos term of the
anomalous $U(1)$ is a function of the moduli.

Nevertheless, our explicit results here show that a scenario similar to this may
not be impossible to realise. The main point is that although moduli are
stabilised at a non-supersymmetric point, the breaking of supersymmetry is
suppressed by inverse powers of the volume or equivalently by powers of $M_s /
M_P$. This makes the first point above approximately correct. The second point
still has to be assumed as in all models of gauge mediation and requires an
explicit realisation. Here the relevant observation is to compare the strength
of gauge mediation $F_X / x$ to the strength of gravity mediation which is
usually taken to be $M_{3/2}$. However as we have seen the proper comparison is
between $F_X / x$ with the size of the gravity mediation soft breaking terms
which are much smaller than the gravitino mass. Regarding the third point an
explicit analysis should be performed in which both the anomalous $U(1)$ and the
moduli are taken into account in the process of moduli stabilisation and
supersymmetry breaking.

Very similar to the recently discussed local F-theory models, we may expand our
model assuming that there exists a source for gauge mediation, which is
parametrised by the vacuum expectation values of a scalar field $\langle
X\rangle = x +\theta^2  F_X$.  This supersymmetry breaking happens in a sector
hidden from the GUT brane and is being mediated by messenger fields, which are
charged under the GUT gauge group. In order not to spoil gauge coupling
unification, this is generically assumed to be a vector-like pair in the
$5+\bar{5}$ representation of $SU(5)$. For our purposes, in this paper we won't
present a viable dynamical stringy realisation of this supersymmetry breaking,
but just assume that there exists an extra sector, which stabilises the new
moduli such that just the field $F_X$ develops a non-zero VEV without spoiling
the LARGE volume minimum for the bulk moduli. This is clearly a strong
assumption, as a dynamical realisation of gauge mediation is known to be
challenging~\cite{Cvetic:2008mh, Dudas:2008qf, Green:2009mx}. We will comment
more on this towards the end of this section.

The gauge mediated gaugino and sfermion masses are of order
\begin{equation}
  M^{\text{gauge}}_{\widetilde{Q}} \sim M^{\text{gauge}}_{\widetilde{G}}
     = \frac{\alpha_X}{4\pi} \, \frac{F_X}{x} \, ,
\end{equation}
where the $\alpha_X / 4\pi$ prefactor is due to the fact that these masses are
induced via a one-loop effect for the gauginos and via a two-loop diagram for
the sfermions. Note that these formulae used a canonical normalised superfield
$X$.

Now, we would like these gauge mediated soft masses to dominate the gravity
mediated ones. In particular, we want the gauge mediated sfermion masses to
dominate over the gravity mediated ones. To get a first impression of the
numerology we get, we also impose the strong constraint that the supersymmetry
breaking $F_X$ already uplifts the negative vacuum energy~\eqref{vacuumenergy}
of the LARGE volume minimum. We therefore require
\begin{equation}
  \frac{F^2_X}{M_P^2} \simeq \frac{M^2_{3/2}}{16 \log
      \left( \frac{M_P}{M_s} \right)} \, \frac{M_s^2}{M_P^2} \; ,
\end{equation}
where $M_s$ is the string scale, leading to the relation
\begin{equation}
  \label{ftermcond}
  F_X \simeq \frac{1}{4 \sqrt{ \log \left( \frac{M_P}{M_s} \right)}} \,
     M_{3/2} \, M_{s} \; .
\end{equation}
Requiring now that $M^{\text{gauge}}_{\widetilde{Q}} > \vert
M^{\text{grav}}_{\widetilde{Q}}\vert$ leads to the moderate bound
\begin{equation}
  \label{xcond}
  x < \frac{\alpha_X}{4\pi} M_P \simeq \unit{10^{16}}{\GeV} \; ,
\end{equation}
where we used the relation~\eqref{sferma}. If there is a further suppression,
i.\,e.\ $M_{\widetilde{Q}} \simeq M_{3/2} / \cV$, then this bound becomes even
more relaxed. For solving the hierarchy problem, one also needs $F_X / x \simeq
\unit{10^5}{\GeV}$. Once one has specified the favourite values for $x$ and
$F_X$, one can use~\eqref{ftermcond} to determine the value of the gravitino
mass, which we would like to stress will be gravity-dominated. Let us discuss
two examples.

\begin{itemize}
  \item In the local F-theory models, it was argued that the best values are
    \begin{equation}
       x \simeq \unit{10^{12}}{\GeV}, \qquad F_X \simeq \unit{10^{17}}{\GeV^2}
    \end{equation}
    which lead to $M_{3/2} \simeq \unit{1}{\TeV}$, which needs a certain amount
    of tuning of $W_0$. However, the light modulus $\tau_b$ has a mass of the
    order
    \begin{equation}
      \label{masslight}
       M_{\tau_b} \simeq M_{3/2} \, \frac{M_s}{M_P} \; ,
    \end{equation}
    which in this case gives $M_{\tau_b} \simeq \unit{1}{\GeV}$. For such a
    light modulus, we expect to face the cosmological modulus problem (CMP).

  \item Let us now require that the light modulus avoids the CMP by having a
    mass $M_{\tau_b} \simeq \unit{100}{\TeV}$. Then according
    to~\eqref{masslight}, the gravity mediated gravitino mass has to be of the
    order  $M_{3/2} \simeq \unit{10^5}{\TeV}$. Using~\eqref{ftermcond}, this
    leads to $F_X \simeq \unit{10^{22}}{\GeV^2}$. For gauge mediated soft masses
    of the order $\unit{500}{\GeV}$, we therefore get $x \simeq \unit{5\cdot
    10^{16}}{\GeV}$, which is slightly beyond the stronger limit~\eqref{xcond}.
    For further suppression of the sfermion masses there is no problem.

  \item In the first case one could ameliorate this problem by allowing
    for a certain tuning of the Higgs mass, so that the supersymmetry breaking
    scale for the visible sector can be larger than $\unit{500}{\GeV}$. Let us
    still have $F_X \simeq \unit{10^{22}}{\GeV^2}$ to avoid the CMP and require
    $x \simeq \unit{5\cdot 10^{14}}{\GeV}$ to satisfy the
    constraint~\eqref{xcond} for gauge mediation dominance. Then the gauge
    mediated soft masses are of the order $\unit{50}{\TeV}$.
\end{itemize}

Finally, let us discuss in which way this simple model of gauge mediation needs
to be improved in order to show that it can really be embedded into string
theory. As we already mentioned, we did not dynamically explain where the SUSY
breaking field $X$ gets its VEV from. Recently, various kinds of models have
been suggested, which, we think, so far are not completely convincing from a
string theory point of view. One promising model is the so-called Fayet--Polonyi
model. It combines an anomalous Peccei--Quinn symmetry with a linear
superpotential in $X$ generated by another D3-instanton wrapping a del Pezzo
surface of size $T_{\text{FP}}$. This gives rise both to a D-term potential with
a $T_{\text{FP}}$ dependent Fayet--Iliopoulos term and an F-term potential form
the linear superpotential. Note, that the latter also depends on
$T_{\text{FP}}$. Now, also taking the K\"ahler potentials into account one has
to show that dynamically really supersymmetry can be broken in such a way that
the desired values for $x$ and $F_X$ arise.\footnote{It was shown
in~\cite{Dudas:2008qf}, that this model with a simple choice of the K\"ahler
potential actually still posses supersymmetric minima.} Moreover, one expects
that also $F_{T_{\text{FP}}} \ne 0$, which gives another source of supersymmetry
breaking. Finally, one has to ensure that the moduli stabilisation in the bulk,
i.\,e.\ of the $\tau_b$ and $\tau_s$ moduli and the moduli stabilisation of the
local $X$ and $T_{\text{FP}}$ moduli decouple. All these challenging questions
are beyond the scope of this paper.

\subsection{Implications for the Cosmological Moduli Problem}

Let us finish this section with some comments about the cosmological moduli
problem~\cite{willy, cosmomoduli, banks}. The cosmological moduli problem refers to the
existence of late decaying moduli, with mass comparable to the gravitino. The
moduli are expected to be displaced from their minimum during the inflationary
epoch, subsequently oscillating about their minimum and red-shifting as matter.
The lifetime of such moduli is $\tau \sim \frac{M_P^2}{m_{\phi}^3} \gg
\unit{1}{\second}$ for $m_{\phi} \lesssim \unit{1}{\TeV}$. Moduli come to
dominate the energy density of the universe, but if they decay too late then
they fail to reheat the universe to temperatures sufficient for nucleosynthesis.
In some ways the moduli problem is the most severe problem facing low-energy
supersymmetry as it is very difficult to construct a viable cosmology with such
long-lived moduli.

There are various possible approaches to this problem. In the absence of
moduli-fixing mechanisms, it may have been hoped that one could stabilise the
moduli at scales far above the gravitino mass. The more that has been learned
about moduli stabilisation the less plausible this scenario has become.

The results in this paper suggest a novel approach to this problem. One of the
properties of local LARGE volume GUTs with D-term stabilisation is that the soft
terms appear at a scale hierarchically smaller than the gravitino mass.
Depending on the extent of cancellations, we have seen that soft terms appear at
an order not larger than $m_{\text{soft}} \sim \frac{M_{3/2}^{3/2}}{M_P^{1/2}}$,
in the case when the dilaton F-term is responsible for uplifting. In all other
cases gaugino masses will be further suppressed, with at least an extra loop
factor as in anomaly mediation, and possibly even as far as $m_{\text{soft}}
\sim \frac{M_{3/2}^2}{M_P}$. For the two extreme cases the gravitino mass
appropriate to \TeV\ soft terms is
\begin{equation}
  m_{\text{soft}} \sim \frac{M_{3/2}^{3/2}}{M_P^{1/2}} \longrightarrow M_{3/2}
     \sim \unit{10^8}{\GeV} \qquad
  m_{\text{soft}} \sim \frac{M_{3/2}^2}{M_P} \longrightarrow M_{3/2}
     \sim \unit{10^{11}}{\GeV} \; .
\end{equation}
Instead of solving the moduli problem by making the moduli heavy and keeping
soft terms comparable to the gravitino mass, this suggests making the gravitino
heavy and having soft terms much lighter than the gravitino mass.

In the LARGE volume models the volume modulus $T_b$ is relatively light and has
a mass $m_{T_b} \sim \frac{M_{3/2}^{3/2}}{M_P^{1/2}}$, while all other moduli
have masses comparable to $M_{3/2}$. In the first case listed above, with a
gravitino mass of around $\unit{10^8}{\GeV}$, the volume modulus has $m \sim
\unit{1}{\TeV}$ and still poses cosmological problems. However in the other
cases $m_{T_b}$ is sufficiently large to decay before nucleosynthesis. In the
case of maximal suppression, with $M_{3/2} \sim \unit{10^{11}}{\GeV}$, then we
have $m_{T_b} \sim \unit{10^7}{\GeV}$ with no cosmological problems. In all
cases the other moduli (for example dilaton and complex structure moduli) have
masses comparable to the gravitino mass and decay very rapidly.

It would also be interesting to study whether these suppressed soft terms would
affect the thermal behaviour of the LARGE volume models studied
in~\cite{Michele}.

\section{Conclusions}

In this paper we have studied the structure of gravity mediated soft terms that
arise when combining LARGE volume moduli stabilisation with local GUTs and
D-term stabilisation of the cycle supporting the GUT brane.

We find that the modulus determining the size of the standard model cycle does
not break supersymmetry and therefore the scale of gravity mediated soft terms
is highly suppressed compared to the gravitino mass.  Both ``standard'' gravity
mediated terms of $\mc{O}(M_{3/2})$ and also known anomaly mediated terms of
$\mc{O}(g^2 M_{3/2}/16 \pi^2)$ vanish. The first non-zero terms appear to arise
at $\mc{O} \left( \frac{M_{3/2}}{\sqrt{\mc{V}}} \right) \simeq
\frac{M_{3/2}^{3/2}}{M_P^{1/2}}$. However it is possible that additional
cancellations occur and suppress the soft terms even further than this down to
$\mc{O} \left( \frac{M_{3/2}^2}{M_P} \right)$. The appearance of these further
cancellations is related to the (in)dependence of the physical Yukawa couplings
on the fields breaking supersymmetry.

The cancellation of contributions to the soft masses of order $M_{3/2}$
introduces several subtleties. In particular, as the soft terms occur at a scale
parametrically smaller than the gravitino mass effects which are normally
negligible become important. We have tried to include all known effects and have
given general arguments as to when cancellations will take place. Nonetheless,
it is important to look for any further possible contributions to soft terms
which could possibly be dangerous. In this respect one would ideally like a
direct stringy computation of soft terms that would bypass the need to go
through the supergravity formalism.

The suppression of soft terms relative to the gravitino mass opens new avenues
for thinking about the cosmological moduli problem. Rather than the traditional
approach of making the moduli heavy while keeping the gravitino and soft terms
around a \TeV, this opens the possibility of having the moduli and gravitino
much heavier than a \TeV\ while still maintaining \TeV scale soft terms.

If the gravitational soft terms are of the order $\frac{M_{3/2}^{3/2}}
{M_P^{1/2}}$, the volume modulus however remains a problem in the LARGE volume
scenario as its mass is much lighter than the gravitino mass and would be the
same order as the soft terms. If further cancellations occur and the soft terms
are of order $\frac{M_{3/2}^2}{M_P}$, then the volume modulus ceases to be a
cosmological problem.

Several scenarios regarding gravity and anomaly mediation are possible and which
of these is actually realised may be model dependent. The main possibilities
are:
\begin{itemize}
  \item{} If the F-term of the dilaton field is responsible for the uplifting to
    de Sitter space, then $F^S \sim \mc{V}^{-3/2}$ and all the soft masses are
    of order $\frac{M_P}{\mc{V}^{3/2}} \sim \frac{M_{3/2}}{\sqrt{\mc{V}}}$. This
    is of the same order as the mass of the lightest modulus, the volume
    modulus, and this field remains dangerous for the cosmological moduli problem.

  \item{} If any other field is responsible for the de Sitter uplifting, the
    dilaton induces gravity mediated gaugino masses of order $\frac{M_P}
    {\mc{V}^2}$ or from anomaly mediation, barring any further cancellation,
    of order $\alpha \frac{M_P}{\mc{V}^{3/2}}$ where $\alpha$ is a loop factor.
    In both of these cases, identifying the gaugino masses with the \TeV\ scale,
    the cosmological moduli problem is absent since the volume modulus would be
    at least as heavy as $\unit{10}{\TeV}$.

  \item{} For each of the two cases of the previous item, gravity mediated
    scalar masses, if not tachyonic, are of order $\frac{M_P}{\mc{V}^{3/2}}$
    and therefore hierarchically heavier than the gaugino masses, indicating a
    minor version of split supersymmetry~\cite{Wells:2003tf, Wells:2004di,
    ArkaniHamed:2004fb}. However if we have perfect sequestering in the sense
    that physical Yukawa couplings do not depend on the K\"ahler moduli fields
    that break supersymmetry, such terms will cancel. However scalar masses will
    always receive a contribution from the dilaton F-term at order
    $\frac{M_P}{\mc{V}^2}$.

  \item{} Since leading order gravity and anomaly mediation contributions to the
    soft terms are suppressed, then other effects have to be considered. In
    particular string loop corrections could be relevant, e.\,g.\ as
    in~\cite{BergHaackKors, BergHaackPajer}, (giving potential contributions to
    scalar masses of order
    $\frac{M_P}{\mc{V}^{5/3}}$~\cite{deAlwis:2008kt, shanta2}) but also a novel
    scenario may be conceived in which the main source of supersymmetry
    breaking for the observable sector is gauge mediation, however the gravitino
    mass remains very large and unlike previous models of gauge mediation, the
    LSP is no longer the gravitino but can be a more standard neutralino.
\end{itemize}

Even though there are several scenarios, we can still extract some general
conclusions from this analysis. First, as emphasised in~\cite{deAlwis:2006nm},
the effects of the de Sitter uplifting play an important r\^{o}le on the soft
breaking terms. This is unlike previous scenarios based on the LARGE volume in
which they were negligible. Second, in all scenarios the gravitino mass is much
heavier than the \TeV\ scale $M_{3/2} \geq \unit{10^{8}}{\GeV}$ which relaxes
the cosmological problems associated to low-energy supersymmetry. Generically
(except in the case that the dilaton is responsible for uplifting) the lightest
modulus is heavier than the soft terms and therefore cosmologically harmless
also.

Finally we point out that even though there are several cancellations that
reduce the value of the volume to have the \TeV\ scale, there is a minimum value
of the volume that can be extracted from this analysis. Namely, the universal
source of gaugino masses due to the dilaton dependence of the gauge kinetic
function, implies that the gaugino masses cannot be smaller than $\frac{M_P}
{\mc{V}^2}$. The same limit appears for scalar masses for the case of perfect
sequestering ($Z=e^{\mc{K}/3}$). This provides a bound for the size of the
overall volume $\mc{V}\sim 10^6-10^7$ in string units which corresponds to a
string scale of order $M_s \sim \unit{10^{15}}{\GeV}$. Combining this with the
recent result~\cite{Conlon:2009xf} that in local models the GUT unification
scale is given by $M_{GUT} \sim M_s \mc{V}^{1/6}$ this gives a unification scale
of the same order as the one expected for supersymmetric GUT models from LEP
precision results of $M_{GUT} \sim \unit{10^{16}}{\GeV}$. If this scenario is
actually realised it would provide an example in which a string model addresses
simultaneously the two positive properties of the MSSM, namely the full
hierarchy problem, without tuning, and obtaining the preferred scale of gauge
unification.

Furthermore, this value of the volume is of the order of magnitude preferred by
models of inflation in order for the inflaton to give rise to density
perturbations of the right amplitude, normalised by COBE\@. In particular a
volume $\mc{V} \sim 10^5-10^7$ was needed to achieve K\"ahler moduli
inflation~\cite{Conlon:2005jm}. It also ameliorates the gravitino mass problem
pointed out in~\cite{Kallosh:2004yh, Conlon:2008cj}.

We consider our results bring closer local string/F-theory models to
honest-to-God string compactifications since we incorporate the main properties
of such models regarding supersymmetry breaking and moduli stabilisation. Many
questions remain open. Concrete examples where the cancellations illustrated
here are realised, including an uplifting term, loop corrections, etc.\ are
desirable. The presence of such sub-leading contributions to soft terms can be
recast in the presence of corrections to the physical Yukawa couplings.
Specifically, the scale of the soft terms can be related to the extent to which
the (local) physical Yukawa couplings depend on the (bulk) supersymmetry
breaking fields. In the limit of perfect sequestering the K\"ahler moduli
contribution to soft masses vanish. It may be possible to study this issue more
precisely using the techniques of orbifold CFT\@. Furthermore, for F-theory
constructions, even though in general they are treated in a way similar to
orientifold constructions, the 4D effective field theory for F-theory models is
less under control. In particular the $\alpha'$-corrections which are crucial in
the large volume scenario, need to be computed for F-theory compactifications.

\subsection*{Acknowledgements}

We gratefully acknowledge discussions with C.\,P.~Burgess, B.~Campbell, K.~Choi,
M.~Cicoli, M.~Dolan, T.~Grimm, L.~Ib\'a\~nez, D.~L\"ust, A.~Maharana,
F.~Marchesano, E.~Palti, E.~Plauschinn, M.~Schmidt-Sommerfeld, A.~Uranga,
G.~Villadoro, T.~Weigand and E.~Witten. FQ~wants to particularly thank S.~de
Alwis for many enlightning discussions on related subjects. RB~would like to
thank the Galileo Galilei Institute for Theoretical Physics for hospitality and
the INFN for partial support during the completion of this work. SLK~would like
to thank the CERN Theory group for hospitality at the early stages of this
project. JC~is grateful to the Royal Society for a University Research
Fellowship.


\appendix

\section{F-Terms}
\label{appa}

As there is a cancellation at leading order taking place in the calculation of
various soft terms, a careful large volume expansion up to the next-to-leading
order has to be performed. Let us start with the expressions for
$e^{-a\tau_s}$ and $\tau_s^{3/2}$ in the minimum, which will be needed later.

For this purpose, consider the scalar potential~\eqref{scapot}. Upon minimising
it with respect to the two independent variables $\tau_s$ and $\cV$, we get two
expression: First, from the condition $\frac{\partial V_F}{\partial \tau_s} =
0$, it follows:
\begin{equation}
  e^{-a\tau_s} = \frac{\mu}{\lambda} \, \frac{|W_0|}{aA \cV} \,
    \frac{1}{\sqrt{\tau_s}} \, \frac{(1-a\tau_s)}{(-2a +
       \frac{1}{2\tau_s})} \; .
\end{equation}
After developing the denominator in powers of $1/(a\tau_s)$ and inserting the
expressions for $\mu$ and $\lambda$ we get
\begin{equation}
  \label{approx1}
  e^{-a\tau_s}  \approx \frac{3}{4}\frac{\eta_s^{3/2}}{aA} \, \sqrt{\tau_s}\,
     \frac{W_0}{\cV}\, \left(1-\frac{3}{4a\tau_s}\right) \; .
\end{equation}
The second expression arises upon solving $\frac{\partial V_F}{\partial \cV} =
0$ for $\tau_s^{3/2}$ and thereby using~\eqref{approx1}. The result is
\begin{equation}
  \label{approx2}
  \tau_s^{3/2} \approx \frac{\hat{\xi}}{2\eta_s^{3/2}} \,
     \left(1+\frac{1}{2a\tau_s} \right) \; .
\end{equation}
Another approximation needed in the following is:
\begin{equation}
    \label{approx3}
    \cK^{a b} \, (\partial_b \cK) = -\frac{4\cV^2+\cV\hat{\xi}
       + 4 \hat{\xi}^2}{2(\cV-\hat{\xi})(\cV+\frac{\hat{\xi}}{2})} \, \tau_a
      \approx -2 \tau_a - \frac{3}{2} \, \hat{\xi} \, \frac{\tau_a}{\cV} \; ,
\end{equation}
where the sum runs only over K\"ahler moduli. The first equality can be derived
using the expressions for the K\"ahler metric and the derivatives of the
K\"ahler potential with respect to the moduli in terms of two-cycle volumes
$t^a$ instead of four-cycle volumes $\tau_a$ (see~\cite{Bobkov:2004cy,
Balasubramanian:2005zx}). We are now in a position to calculate $F^b$:
\begin{equation}
  F^b = e^{\cK/2} \, \cK^{bJ} D_J W = e^{\cK/2}\, \left( \cK^{b\tau_j} \,
      (\partial_{\tau_j} \cK) W + \cK^{bs} \, (\partial_s W)
       + \cK^{bS}D_S W\right) \; .
\end{equation}
The term involving $D_S W$ turns out to be sub-leading in the $\cV^{-1}$
expansion and can be neglected (see below). The derivative of the superpotential
with respect to $T_s$ undergoes a sign-flip due to the minimisation with respect
to the corresponding axion as argued in~\cite{Balasubramanian:2005zx}. Using the
approximations~\eqref{approx1},~\eqref{approx2} and~\eqref{approx3}, one easily
gets:
\begin{equation}
 F^b = -2\tau_b \frac{\sqrt{g_s}}{\sqrt{2}} \, \frac{W_0}{\cV}
     - \frac{3}{8\sqrt{2}} \, \frac{\tau_b}{a\tau_s} \,
   \left(1+\frac{3}{2a\tau_s}\right)\, \frac{W_0}{\cV^2}
   + \cO(\cV^{-3}) \; ,
\end{equation}
or with the expressions for the gravitino- and gaugino-mass~\eqref{masses_grav}
and~\eqref{gaugino_mass} inserted:
\begin{equation}
 F^b = -2\tau_b M_{3/2} - \frac{\tau_b}{2a\tau_s} \,
   \left(1+\frac{3}{2a\tau_s}\right)\, M_{\widetilde{G}}
   + \cO(\cV^{-3}) \; .
\end{equation}
From~\eqref{approx1}, it can be derived that $a\tau_s \approx \ln \cV \approx
10$. Thus, for the sake of shorter formulae, one may also neglect the second
term in the parenthesis:
\begin{equation}
  \boxed{
 F^b \approx -2\tau_b M_{3/2} - \frac{\tau_b}{2a\tau_s} \,
   M_{\widetilde{G}}
  }
\end{equation}

Next, we want to calculate $F^S = e^{\cK/2} K^{S\bar{J}}D_{\bar{J}} \bar{W}$. Here,
a subtlety arises concerning $D_S W = \partial_S W + W (\partial_S \cK)$: the
K\"ahler potential depends on the dilaton not only in the usual way via $-\ln
(S+\bar{S})$, but there is also a contribution in the $\alpha'$-correction in the
K\"ahler moduli part. Thus, $\partial_S \cK$ has $\cV^{-1}$ corrections:
\begin{equation}
  \label{dilaton-fterm}
  D_S W\approx \partial_S W_0 - \frac{g_s}{2}W_0
     - \frac{3}{4}\frac{\xi}{g_s^{1/2}} \frac{W_0}{\cV} + \cO(\cV^{-2} ) \; .
\end{equation}
Also as a consequence of the $\alpha'$-corrections, the minimum of the scalar
potential for the dilaton is shifted away from the supersymmetric locus $D_S W=
0$ at order $\cV^{-1}$. In order to determine the new minimum, one would have to
minimise the full potential, before integrating out the dilaton. However, since
we do not have an explicit model with a full flux sector, in order to capture
this effect, we assume that the two leading order terms in~\eqref{dilaton-fterm}
cancel and keep only the next-to-leading order terms in the $\cV^{-1}$
expansion. The expression we get in this way has certainly the correct order in
$\cV$ and we include an order one constant $\gamma$ in the final result
comprising the uncertainty about the true location of the new minimum.
\begin{equation}
  D_S W\approx  - \frac{3}{4}\gamma' \frac{\xi}{g_s^{1/2}} \frac{W_0}{\cV} \; .
\end{equation}
In the sum over $D_I W$ in the dilaton F-Term $F^S = e^{\cK / 2} \cK^{SJ} D_J
W$, there are finally two contributions at order $\cV^{-2}$: one from $\cK^{Sb}
F_b$ and one from $\cK^{SS} F_S$. The result reads:
\begin{equation}
  \boxed{
    F^S \approx  \frac{3}{2\sqrt{2}}\gamma\frac{\xi}{g_s^2} \frac{W_0}{\cV^2}
  }
\end{equation}

\section{Vanishing D-terms including matter}

We will consider in this appendix a concrete example with a generic D-term
including not only the field dependent FI-term but also a charged matter field.
In general vanishing D-terms do not imply vanishing FI-term but a cancellation
between the two terms entering the D-term potential. We argue here
(following~\cite{Cascales:2003wn}) that once soft supersymmetry breaking terms
are included, as long as the square of the scalar masses is positive the minimum
of the scalar potential is for vanishing both matter field VEV and FI-term.

Since in local models the standard model cycle is a del Pezzo surface that can
and usually prefers to shrink to small size, it is dangerous to work in the
regime where the cycle size is larger than the string scale. Even though at
sizes of the order of the string scale the spectrum and couplings of the model
are not understood, the regime close to a del Pezzo singularity is under a much
better control, the spectrum is determined by the extended quiver diagrams and
the low-energy effective theory can be reliably used in an expansion in the
small blow-up mode.

This effective field theory has been recently discussed in~\cite{Conlon:2008wa}.
We start with the same background geometry as before including one large
$\tau_1$ and two small cycles $\tau_2$, $\tau_3$. On the rigid cycle $\tau_2$ we
have the standard non-perturbative effect. Being at the singular locus for
$\tau_3$, the effective field theory can be approximated by the following
supergravity set-up:
\begin{equation}
  \begin{split}
    \mc{K} &= -2 \log{ (\cV + \tfrac{\hat{\xi}}{2} ) } +\frac{\alpha \tau_3^2}{\cV}
      + Z |\varphi|^2 \; , \\
         W &= W_0 + A e^{-a T_2} \; , \\
         f &=  d\, T_3 + S \; ,
  \end{split}
\end{equation}
where $\varphi$ denotes a matter field that is charged under an anomalous $U(1)$
on the standard model cycle, as is the cycle volume itself. As discussed
in~\cite{Conlon:2008wa}, the effective theory for $\tau_3$ differs from the
standard treatment for relatively large values of $\tau_3$ since we are working
close to the singularity. The anomalous $U(1)$ generates a D-term potential with
a Fayet--Iliopoulos term:
\begin{equation}
  V_D = \frac{1}{2(d \tau_3+s)} \left( Q_\varphi Z |\varphi|^2
     + \frac{Q_{\tau_3}\tau_{3}}{\mc{V}} \right)^2 \; .
\end{equation}
The matter metric $Z$ is taken to have the general form
\begin{equation}
  Z = \frac{1}{\cV^{2/3}} \left(\beta + \gamma \tau_3^\lambda
         - \frac{\delta}{\cV} \right) \; ,
\end{equation}
where the constants $\beta$, $\delta$ can in principle depend on the dilaton and
complex structure moduli.

The D-term potential determines the size of $\tau_3$ and implies
\begin{equation}
  \tau_3 \sim |\varphi|^2 \cV^{1/3} \; .
\end{equation}
For a vanishing VEV of $\varphi$ this implies as previously $\tau_3 = 0$.
Expanding around $\varphi=0$, the scalar potential is given by the standard
LARGE volume potential and at next-to-leading order by a contribution quadratic
in $\varphi$:
\begin{equation}
  \begin{split}
    V &= \frac{1}{(\cV+\tfrac{\xi}{2})^2} \left( \frac{8}{3}|a A|^2
         \sqrt{\tau_2} \tau_1^{3/2} e^{-2a\tau_2} - 4 W_0 a A \tau_2
         e^{-a\tau_2} + \frac{W_0^2 3\xi}{4 \tau_1^{3/2}} + Y \right. \\
      &- \left. \frac{\beta|\varphi|^2}{3 \tau_1} \left(\frac{8}{3}|a A|^2
         \sqrt{\tau_2} \tau_1^{3/2} e^{-2a\tau_2}  -4 W_0 a A \tau_2
         e^{-a\tau_2} + \frac{9 W_0^2 (5\frac{\delta}{\beta} + 2\xi)}
         {4 \tau_1^{3/2}} \right) \right) \\
      &+ \frac{\beta |\varphi|^2}{\tau_1(\cV+\tfrac{\xi}{2})^2} \left(
         \frac{8}{3}|a A|^2 \sqrt{\tau_2} \tau_1^{3/2} e^{-2a\tau_2}
         -4 W_0 a A \tau_2 e^{-a\tau_2} + \frac{3 W_0^2 \xi}{4 \tau_1^{3/2}}
         +Y \right) \; ,
  \end{split}
\end{equation}
where the last term arises from the expansion of $e^{\mc{K}}$ and $Y$ denotes
the F-term uplifting term, which allows for a stabilisation at zero vacuum
energy.

With zero vacuum energy, the mass of $\varphi$ is given by
\begin{equation}
  \begin{split}
    m_{\varphi}^2 &= {\mc{K}}^{-1}_{\varphi\varphi} \frac{-\beta}{3 \tau_1^4}
         \left( \frac{8}{3}|a A|^2 \sqrt{\tau_2} \tau_1^{3/2} e^{-2a\tau_2}
          - 4 W_0 a A \tau_2 e^{-a\tau_2} + \frac{9 W_0^2 \left( 2\xi
          - 5\frac{\delta}{\beta}\right)}{4 \tau_1^{3/2}}\right) \\
                  &= -\frac{1}{3 \tau_1^3} \left(V_{\text{min}}
          + \frac{45 W_0^2 \left(\frac{\xi}{3} - \frac{\delta}{\beta}\right)}
             {4 \tau_1^{3/2}}\right)
          \approx \frac{15 W_0^2 (\frac{\delta}{\beta} - \frac{\xi}{3})}
             {4 \tau_1^{9/2}} \, .
  \end{split}
\end{equation}
Different ratios of $\delta / \beta$ allow for tachyonic, zero or positive
masses at this order. In particular:
\begin{equation}
  \frac{\delta}{\beta}
    \begin{cases}
      < \frac{\xi}{3} & \text{tachyonic,} \\
      = \frac{\xi}{3} & \text{zero,}\\
      > \frac{\xi}{3} & \text{positive.}
    \end{cases}
\end{equation}
With respect to the matter metric the condition $\frac{\delta}{\beta} =
\frac{\xi}{3}$ can be understood as follows: The case of vanishing masses
corresponds to the following matter metric:
\begin{equation}
  Z = \frac{\beta}{\cV^{2/3}} \left(1 - \frac{\xi}{3 \cV} \right)
     \approx \frac{\beta}{(\cV + \tfrac{\xi}{2})^{2/3}}
     = \beta e^{\mc{K}/3} \; ,
\end{equation}
which is the condition found in section~\ref{sec:decoup} for extreme
sequestering and cancellation of scalar masses at the $1/{\mc{V}}^{3/2}$ level.
Without the uplifting term, the effect of the term arising from the expansion of
$e^{\mc{K}}$ is generally sub-leading to the other contribution since it is
suppressed with $1/a\tau_2$.

For positive scalar masses we can clearly see that combining the term
$m_\varphi^2 \varphi^2$ with the D-term potential, both the VEV of $\varphi$ and
the FI-term vanish at the minimum as desired. For the tachyonic case this would
indicate as usual that at the minimum the scalar field and the FI-term would be
non-vanishing. If $\varphi$ is a field charged under the standard model gauge
group this is undesirable since it would break the standard model symmetries at
high energies. If the condition $\frac{\delta}{\beta} = \frac{\xi}{3}$ is
satisfied the positivity of the squared scalar masses at a higher order would
have to be determined.

\clearpage
\bibliography{rev}
\bibliographystyle{JHEP}


\end{document}